\documentclass[10pt,a4paper,dvips,twocolumn,showpacs]{revtex4}
\usepackage[english]{babel}
\usepackage{epsfig}
\usepackage{graphics, curves} 
\usepackage{amsmath, amsfonts, amssymb}
\usepackage{graphicx}
\usepackage{psfrag}
\usepackage{dcolumn}
\usepackage{dcolumn}

\newcommand{\Keffx}{F_x}
\newcommand{\Keffy}{F_y}
\begin{document}
\title{Permeability Estimates of Self-Affine Fracture Faults Based on Generalization of the Bottle Neck Concept}

\author{Laurent Talon}
\email[talon@fast.u-psud.fr]{}
\affiliation{Univ.\ Pierre et Marie Curie-Paris6, Univ.\ Paris-Sud, CNRS,
Lab.\ FAST, B{\^a}t.\ 502, Campus Univ., Orsay, F--91405, France.}

\author{Harold Auradou}
\email[auradou@fast.u-psud.fr]{}
\affiliation{Univ.\ Pierre et Marie Curie-Paris6, Univ.\ Paris-Sud, CNRS,
Lab.\ FAST, B{\^a}t.\ 502, Campus Univ., Orsay, F--91405, France.}

\author{Alex Hansen}
\email[Alex.Hansen@ntnu.no]{}
\affiliation{Department of Physics, Norwegian University of Science and
Technology, N--7491 Trondheim, Norway}

\date{\today}

\begin{abstract}
We propose a method for calculating the effective permeability of
two-dimensional self-affine permeability fields based on generalizing the
one-dimensional concept of a bottleneck. We test the method on fracture
faults where the local permeability field is given by the cube of the
aperture field.  The method remains accurate even when there is
substantial mechanical overlap between the two fracture surfaces. 
The computational efficiency of the method is comparable to calculating a
simple average and is more than two orders of magnitude faster than
solving the Reynolds equations using a finite-difference scheme.
\end{abstract}

\maketitle
In many low permeability geological formations, flow occurs primarily through 
fracture networks \cite[]{NAS}.  In order to model such systems and to predict their 
behavior, there is a 
need for reliable modeling of the hydromechanical behavior of fracture.
We consider in this note the situation where the shear displacement between 
the fracture walls strongly affects its permeability. Because of its 
relevance, this situation has been considered in many recent hydromechanical 
studies 
\cite[]{archambault1997,yeo1998,hans2003,auradou2005,matsuki2006,watanabe2008,nemoto2009}.
Laboratory tests report that the shearing process results in 
a significant channelization of the flow and an enhancement of the 
permeability in the direction normal to the shear. This behavior is found to 
be related to the long-range spatial organization of the void space, and 
efforts have been undertaken to modelize such 
system in order to provide upscaled value for the fracture permeability.
Recently Mallikanas and Rajaram \cite[]{mallikamas2005} 
determined analytically 
using perturbation analysis of the Reynolds equation 
to the lowest non-trivial order the fracture permeability. This model,
however, does not take into account the role of contact areas and will fail
if they appear.  The effect of contacts may, however, be taken into account 
by introducing an empirical parameter \cite[]{zimmerman1996} that is strongly influenced
by the number and the spatial distribution of the contacts \cite[]{li2008}.  

\begin{figure}
\includegraphics[width=18pc,clip]{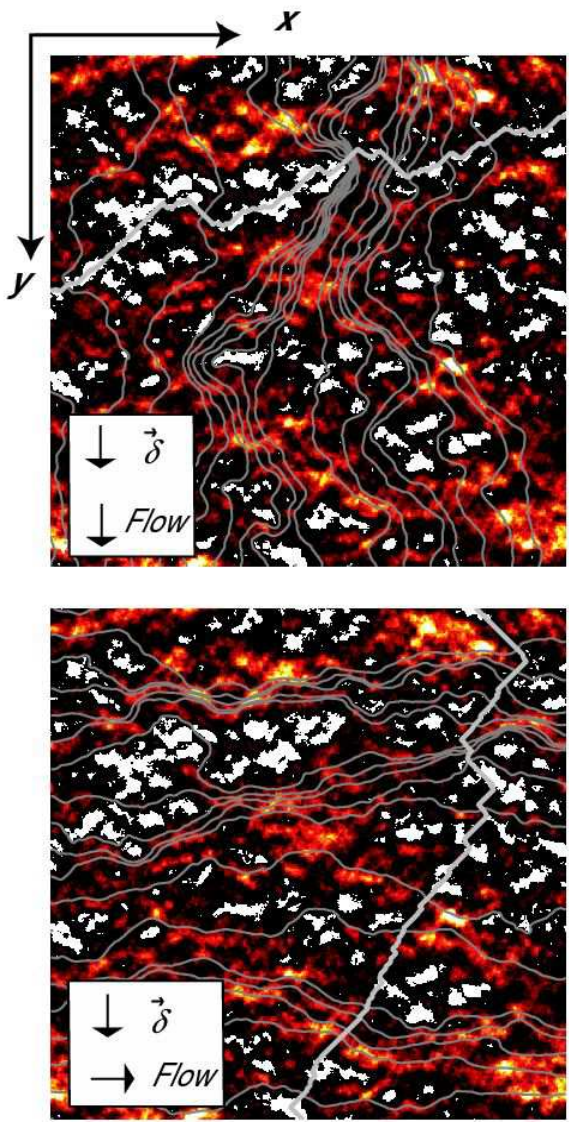}
\caption{ \label{fig1}  Aperture field obtained by shifting laterally
by $\vec{\lambda}=10\vec{e_y}$ two matching self-affine surfaces with Hurst
exponent $\zeta=0.8$. The size is $512\times 512$ and the
mean aperture is $\langle h\rangle=7$.  Darker shades mean smaller
aperture whereas lighter shades means larger apertures.  White zones
are contact areas.  Top (resp.\ bottom) figure: the flow is
normal to (resp.\ along) the lateral displacement $\vec{\delta}$.  
The flow lines are shown as grey paths.  The worst paths normal to the 
average flow directions in the two cases are shown as thick grey lines.
In the top figure, we have 
$\langle h^3\rangle/\int_{\cal C} d\vec\ell\cdot {\vec e}_y h(\vec \ell)^3=3.32$ and
in the bottom figure 
$\langle h^3\rangle/\int_{\cal C} d\vec\ell\cdot {\vec e}_x h(\vec \ell)^3=8.54$.}
\end{figure}

We present in this note a computational method for calculating the
permeability of such fracture faults even in the presence of contacts.  
This new method scales linearly with the number of grid points and is more 
than two orders of magnitude faster than solving the finite-differenced Reynolds 
equations through LU decomposition.  

There is now ample experimental and observational evidence that fracture
surfaces are self affine, see e.g., 
\cite[]{mpp84,brown1985,ptbbs87,blp90,mhhr92,sgr93,pkhrs95}.
A self-affine fracture may characterized by a rescaling $r$ of distances
in the average fracture plane and a rescaling $r^\zeta$ of distances in
the orthogonal direction leaves the statistical properties of the surface
unchanged.  Here $\zeta$ is the Hurst exponent.  We use in the following
$\zeta=0.8$, i.e.\ the value often
reported for rocks fractures \cite[]{poon1992,sgr93,matsuki2006}.  
When the two matching fracture surfaces are displaced by a distance $\lambda$ along
the average fracture plane, the ensuing aperture field will be self-affine
up to the length scales of the order of $\lambda$.  On larger scales, the aperture
field settles to a constant value proportional to the average fracture
opening \cite[]{pkhrs95}.  This gives rise to a aperture field $h=h(x,y)$.

As the two fracture surfaces approach each other, they will eventually 
come into contact and hence overlap.  Overlap also occurs if the gap 
between the two fracture surfaces remains fixed while the lateral 
displacement $\lambda$ increases. At such places of contact, we set
the aperture $h(x,y)$  to zero.  The contact areas are
shown as white in Fig.\ \ref{fig1}.  
The lateral displacement results in a strong structural anisotropy:
The contact zones are more elongated in the direction normal
to the displacement ($x$ direction with reference to Fig.\ \ref{fig1})
than in the other direction ($y$ direction with reference to Fig.\ \ref{fig1}).

The idea behind the method we introduce in this note is based on the 
generalization of the concept of the bottle neck to higher dimensions.
Some forty years ago, \cite[]{ahl71} presented a different generalization
of the same concept.  As we shall see, only in a limiting case do the two
generalizations approach each other.   

Before delving into the two-dimensional generalization --- and hence the
method we present --- we discuss the one-dimensional case.  Hence we
have a channel with self-affine walls that have been translated relative
to each other along the direction of the channel by a distance $\lambda$. The
channel aperture is given by $h=h(y)$, where the $y$ axis is oriented along
the channel.  Assuming that the Reynolds equations govern the flow in the 
channel, the local permeability is proportional to $h(y)^3$. The permeability
of the entire channel is then given by the harmonic mean of the local
permeability, i.e.,  $\propto \langle h(y)^{-3} \rangle^{-1}$ 
\cite[]{zimmerman1996}.
If we now assume that the two channel walls are brought close together 
(so that $\langle h(y)\rangle$ decreases), the permeability is increasingly 
controlled by the region of minimum aperture
$\min_y h(y)^3$ \cite[]{gh95,shg99}, which may be then viewed as a bottle neck.

How wide, $\Delta$, is the bottle neck region?  This of course depends on the
geometry of the two channel walls in this region.  For
the time being, we leave $\Delta$ as a parameter. We
now divide the entire channel along the $y$ axis into two regions: The
bottle neck region which has a width $\Delta$ and the rest which has a 
width $L-\Delta$, where $L$ is the length of the entire channel.   
The bottle neck region have a permeability essentially given by 
$\min_y h(y)^3/\Delta$ and the rest of the channel will have a
permeability that is essentially $\langle h^3(y)\rangle/(L-\Delta)$.
The total permeability of the channel $K_y$ may then be approximated by 
$D_y$ given by
\begin{equation}
\label{totalperm-x}
\frac{L}{D_y}=
\frac{\Delta}{\min_{y}h(y)^3}+\frac{L-\Delta}{\langle h(y)^3\rangle}\;,
\end{equation}
Clearly, $\Delta$ will evolve as the average channel width 
$\langle h(y)\rangle$ decreases and keeping it constant will constitute an
approximation.  How good is such an approximation? A natural choice for a 
fixed $\Delta$ may e.g.\ be the discretization length scale (i.e., the 
lattice constant).  As $\langle h(y)\rangle$ decreases, the more dominant 
the bottle neck region will be and the more sensitive $D_y$ will be to the 
discretization at this point. Approximations are unavoidable as the average 
channel width decreases.  A ``natural" choice as the discretization length 
itself breaks down when the discretization itself breaks down.  

We now turn to generalizing this discussion to two-dimensional aperture 
fields $h=h(x,y)$. The main difference between the one-dimensional channel 
and the two-dimensional fracture is that flow can in the latter case easily 
bypass regions of small aperture.  They do not play the crucial role here as 
they did in the one-dimensional channel.

We therefore generalize the concept of the bottle neck for two-dimensional
fractures.  As a first step to this generalization, we consider 
paths going from one side of the fracture to the opposite side cutting
across the average flow direction.
As a result of mass conservation, the flow has to pass through all such
paths.  For each transverse path $\cal C$ with respect to the flow direction (here, the 
$y$ direction), we may calculate the average aperture 
cubed along it,
\begin{equation}
\label{effh3}
L_{\cal C}
\langle h^3\rangle_{\cal C}=\int_{\cal C} d\vec 
\ell\cdot {\vec e}_x h(\vec \ell)^3\;,
\end{equation}
where $L_{\cal C}$ is the length of the path and ${\vec e}_x$ is the unit
vector in the $x$ direction.  This average now replaces for the two dimensional
system, the local permeability $h^3(y)$ for the channel in one 
dimension.  

In the one-dimensional channel we then went on to identifying the smallest
local permeability $\min_y h^3(y)$.  This was the bottle neck.  In two
dimensions, we now search for the {\it path with the smallest average
aperture cubed,\/} henceforth refered to as the {\it worst path,\/}
\begin{equation}
\label{smallesth3}
\min_y h(y)^3\to \min_{\cal C}\int_{\cal C} d\vec \ell\cdot {\vec e}_x
h(\vec \ell)^3\;.
\end{equation}
Fig.\ \ref{fig1} shows for one of the realizations the two
worst paths obtained for flow directions along and normal to the
lateral displacement.

Using the same reasoning as in one dimension, we may now generalize 
Eq.\ (\ref{totalperm-x}) by replacing
$\min_y h(y)^3$ by $ \min_{\cal C}\int_{\cal C} d\vec \ell\cdot 
{\vec e}_x h(\vec \ell)^3$, hence
\begin{equation}
\label{totalperm}
\frac{L}{W D_y}=
\frac{\Delta}{\min_{\cal C}\int_{\cal C} d{\vec\ell}\cdot 
{\vec e}_x h(\vec\ell )^3}
+\frac{L-\Delta}{W \langle h(x,y)^3\rangle}\;,
\end{equation}
where $W$ is the width of the fracture in the $x$ direction.

When the average flow is in the
$x$ direction rather than the $y$ direction, there is
of course an equivalent expression for $D_x$.
These two expressions, for $D_y$ and $D_x$,
form the core of our method for approximating the permeability of
fracture faults.  

We now discuss briefly the relation between the worst path method and 
the Ambegaokar-Halperin-Langer estimate \cite[]{ahl71}. The AHL estimate is 
based on the idea that when the local permeability is very widely distributed,
the upscaled permeability
is controlled by the smallest local permeability along the path connecting the
inlet to the outlet that has the highest average permeability along 
it.  If in Eq.\ (\ref{effh3}), we assume that $h^3(x,y)$ is so
widely distributed that the integral is dominated by the {\it largest
value\/} of $h^3(x,y)$ along the path, the integral becomes
\begin{equation}
\label{ahleffh3}
\langle h^3\rangle_{\cal C} = \max_{\vec \ell \in \cal C}  
h(\vec \ell)^3\;.
\end{equation}
If we now combine this expression with Eq.\ (\ref{smallesth3}) to estimate
the permeability of the bottle neck region, we find
\begin{equation}
\label{ahlsmallesth3}
\min_{\cal C}\left[\max_{\vec\ell\in\cal C} 
h(\vec \ell)^3\right]\;.
\end{equation}
This expression is essentially the Ambegaokar-Halperin-Langer expression for
the permeability, except that we in this limit end up with the maximum
permeability along the path with the minimum permeability along it, whereas
in the analysis of \cite[]{ahl71}, ``min" and ``max" have been substituted.
In two-dimensional systems, this is equivalent.  Hence,
only in the limit of extremely broad aperture distributions, is our 
formulation equivalent to that of \cite[]{ahl71}.   

As in one dimension, the width of the bottle neck region,
$\Delta$, is a parameter depending on the local topography near the 
worst path.  It needs to the determined independently.
One way to estimate it is to equate $D_y$ (resp.\ $D_x$),
gotten from Eq.\ (\ref{totalperm}), with the permeability gotten from another 
method when the fracture opening is large: the detail of the procedure 
is described farther in the text.  As in one dimension, we expect 
$\Delta$ to change as the average fracture aperture, $\langle h(x,y)\rangle$
is lowered.  Assuming that it is a constant --- as we will do ---
constitutes an approximation.

Given the aperture fields, we compared their permeabilities found using Eq.\
(\ref{totalperm}) with the results of two other techniques.  The
first one, proposed by \cite[]{gelhar1983}, is based on a stochastic
continuum theory applied to a first order perturbation expansion of
the Darcy's law. We compute numerically the Fourier transform of 
the permeability field perturbation 
$\hat{K(k_x,k_y)} = FT(h^3(x,y) - \langle h(x,y)^3 \rangle )$. 
The two component of the
effective permeability are then calculated from the integrals
\begin{equation}
\label{effkx}
\frac{\Keffx}{\langle h^3\rangle} =  
1 -  \iint \frac{k_x^2}{k^2} 
\frac{|\hat{K}|^2}{\langle h^3\rangle^2} dk_x\ dk_y\;,
\end{equation}
and
\begin{equation}
\label{effky}
\frac{\Keffy}{\langle h^3\rangle} = 
1 -  \iint \frac{k_y^2}{k^2} 
\frac{|\hat K|^2}{\langle h^3\rangle^2} dk_x\ dk_y\;.
\end{equation}
Since this is only a second order expansion, these results are expected
to be valid only for small permeability fluctuations, i.e., when the
fracture opening $\langle h\rangle$ is large compared to the height
fluctuations in the fracture \cite[]{mallikamas2005}. The second method consists in 
solving the flow field inside the permeability field by using a lattice 
Boltzmann method. In the this scheme, we introduce a body force to produce a
Darcy-Brinkman equation as described in
\cite[]{martys2001,talon2003}. We decrease the Brinkman term so
that it has no appreciable effect on the permeability. When the
two surfaces are in contact, the lattice site is set to be solid by
using the ``bounce-back'' reflection method for the density
distribution. A pressure-imposed boundary condition is used at the
inlet and outlet as described by \cite[]{zou1997}.

Practically, we identify the worst path and the corresponding integral, Eq.\
(\ref{smallesth3})
by using a transfer matrix algorithm \cite[]{bs95}. If the average
flow direction is in the $y$ direction, the path we construct
runs between the sides of the sample parallel to the $x$ axis.
We discretize the aperture field $h(x,y) \to
h(i,j)$, onto a square lattice where $i$ runs from 1 to $M=W/a$ and $j$ 
from 1 to $N=L/a$, and $a$ is the lattice constant. We introduce a second field $p(i,j)$
which initially is set to zero everywhere.  We then update layer by layer
in the $i$ direction
\begin{eqnarray}
\label{layer}
p(i+1,j)=&\nonumber\\
\min[p(i,j-1), p(i,j), p(i,j+1)]+h(i+1,j)^3\;,
\end{eqnarray}
until $i=M-1$.  The integral Eq.\ (\ref{smallesth3}) is then given by
\begin{equation}
\label{wp}
p(M,j_M)=\min_{j} p(M,j)\;.
\end{equation}
where we designate by $j_M$
the $j$ value where the mininum $p$ was identified.
In order to reconstruct the worst path, we start at the position
$(M,j_M)$. We then move on the next layer, and identify
$\min[p(M-1,j_M-1),p(M-1,j_M),p(M-1,j_M+1)]$.  The $j$ value
that corresponds to the minimum at level $i=M-1$ is designated $j_{M-1}$.
We then repeat this algorithm until we have identified $j_1$.  The sequence
$j_i$ where $i=1,\cdots,M$ gives the coordinates of the worst path.

\begin{figure}
\includegraphics[width=18pc,clip]{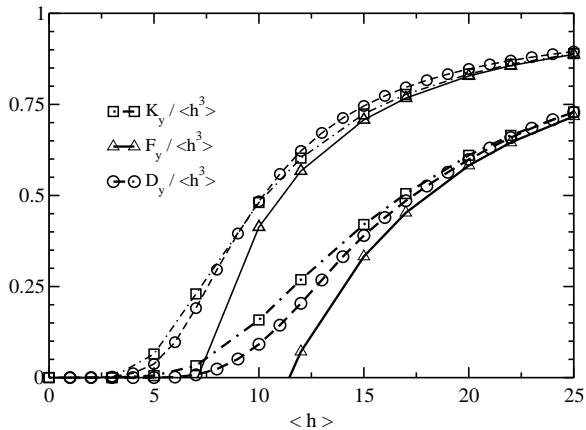}
\caption{\label{fig2} Normalized permeabilities of a rough fracture for
flow along the lateral displacement as function of the
fracture opening $\langle h\rangle$ for two lateral displacements
(Thick solid curves $\vec{\lambda}=10\vec{e_y}$, solid curves 
$\vec{\lambda}=20\vec{e_y}$).
Circles: permeability, $D_y$,
calculated using Eq.\ (\ref{totalperm}). Squares and triangles are for the
permeabilities obtained by the Lattice Boltzmann ($K_y$) algorithm and by
the second-order perturbation theory ($F_y$). The system size
is $512\times 512$ and $\Delta$ has been set equal to $33$ for
$\lambda=10$ and $38$ for $\lambda=20$ by matching $D_y$ and $F_y$ 
for the maximum fracture opening.}
\end{figure}

\begin{figure}
\label{fig3} \noindent\includegraphics[width=18pc,clip]{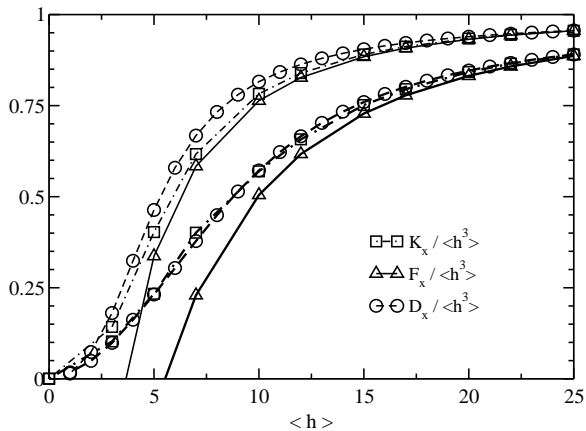}
\caption{Normalized permeabilities of a rough fracture for
flow normal to $\vec{\lambda}$ as function of the
fracture opening $\langle h\rangle$ for two lateral displacements
(Thick solid lines $\vec{\lambda}=10\vec{e_y}$ -
Solid lines $\vec{\lambda}=20\vec{e_y}$). Despite the flow
direction, the conditions are similar to the ones of Fig. \ref{fig2}
and circles, squares and triangles referred
to $D_x$, $K_x$ and $F_x$.}
\end{figure}

In Eq.\ (\ref{layer}) we are assuming that the paths only connect 
nearest-neighbor and next-nearest-neighbor nodes on the lattice, i.e., 
$(i,j\pm k)$ with $(i+1,j)$ where $k=0,1$.  This may be generalized to 
$k=0,\cdots,m$.  In our numerical calculations presented in Figs.\ \ref{fig2} 
and \ref{fig3} we have used $m=2$.  However, we see no appreciable 
difference between this value and
$m=1$.

The algorithm described in Eq.\ (\ref{layer}) assumes that the paths
do not turn back, i.e., $j_c=j_c(i)$.  In very strongly disordered fractures,
such turns may play a role.  This is not the case for the fractures studied 
here.
However, when turns do appear, different and more
involved algorithms must be used \cite[]{hh92,hk04}. Whereas the
algorithm described in Eq.\ (\ref{layer}) scales as the number of
nodes $M\times N$ in the discretized height field, the algorithms
capable of handling overhangs scales as $M^2\times N^2$.

We first study the situation where flow is parallel to the lateral 
displacement, i.e.\ orthogonal to the
channelization. Such flow situation is illustrated in top figure in Fig.\
\ref{fig1}. Fig.\ \ref{fig2} shows the variation of the permeability of
the fracture estimated by the lattice-Boltzmann method (squares) and by the
second order expansion (triangles) as functions of the mean fracture opening
and for two lateral displacements.  For each of the aperture fields, $p(M,j_M)$
as well as $\langle h^3\rangle$ were measured.  The estimation of $D_y$ still
requires an estimation of $\Delta$, see Eq.\ (\ref{totalperm}).  For the two
lateral displacements and for large fracture openings, the second order 
estimate of the permeability, $F_y$, Eq.\ (\ref{effky}) fits
the lattice-Boltzmann $K_y$ well.  In this region, we equate $F_y$ and $D_y$,
hence determining $\Delta$.  We then go on to using {\it the same\/}
$\Delta$ for all subsequent fracture openings.  Herein lies the major
approximation in our method.  
As soon as contact areas appear (here a noticeable difference
occurs when contacts cover about $10\%$ of the total fracture area),
the perturbative estimate $F_y$ fails to describe the continuous
drop of the permeability whereas $D_y$ remains very close to $K_y$.

\begin{figure}
\label{fig4} \noindent\includegraphics[width=18pc,clip]{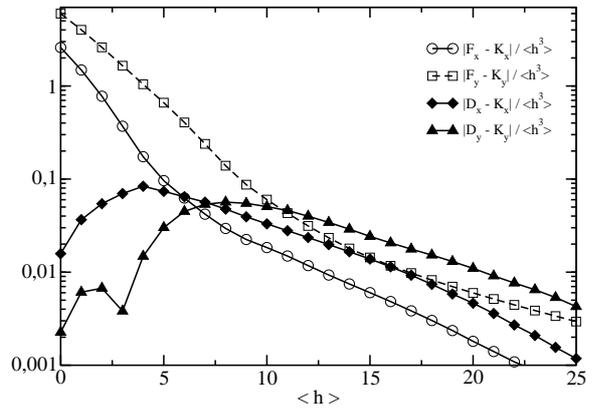}
\caption{The relative errors $|F_x-K_x|/\langle h^3\rangle$,
$|F_y-K_y|/\langle h^3\rangle$, $|D_x-K_x|/\langle h^3\rangle$, and 
$|D_y-K_y|/\langle h^3\rangle$ as a function of the average fracture
aperture $\langle h\rangle$.  The data has been averaged over 10 samples,
each of size $512 \times 512$, Hurst exponent $H=0.8$ and lateral 
displacement $u=10$.}
\end{figure}

For flow normal to the
lateral displacement and, as illustrated in the bottom figure of
Fig. \ref{fig1}, we find strong channelization and it is markly different
from the one observed when flow is parallel to $\vec{\lambda}$. Fig.\
\ref{fig3} shows the permeabilities found by the three
methods: a drop off of the permeability with the fracture closure is
observed but is less marked than for flow along the shift (See Fig.\
\ref{fig2} for comparison). As previously mentioned, as soon as
contacts between the two surfaces occur the pertubative method fails
to describe the marked permeability decrease observed with the
lattice boltzmann method. Yet the
worst path method still accurately captures
the permeability reduction estimate in the direction parallel to the
channelization even for large lateral displacement of the fracture
walls.

We show in Fig.\ \ref{fig4}, the relative errors between the
worst path method and the lattice Boltzmann method, and the perturbative
approximation and the lattice Boltzmann method for different average
fracture openings.  The data have been averaged over ten samples.  As we
see, the worst path method performs very well for all values of the 
average fracture opening and for the two flow directions.

To conclude, we have introduced a new technique to estimate the
permeability of self-affine fracture faults.  Compared to other approximative 
methods, it performs very well by being able to reproduce the permeability 
closely even when the fracture opening tends to zero.
``Exact" methods such as the lattice Boltzmann method gives more precise
results.  However, the computation time is reduced by several orders of 
magnitude compared to alternative methods. To our knowledge,
solving the finite-differenced Reynolds equations through LU decomposition
is the fastest ``exact method".  For the samples studied in this note, the
worst path method used 0.01 seconds per sample and per average fracture
opening, whereas the LU decomposition used from 3 to 8 seconds.  Both methods
scale linearly with the number of nodes.

A length scale $\Delta$ is introduced in order to fit
the permeability measured. This length characterizes the extension in the flow
direction of the region dominated by the worst path.  We assume that $\Delta$
remains constant as the average fracture opening is changed.  This is one
of the major approximation build into the method --- but it allows us to
determine $\Delta$ by comparison with other approximate methods such as the 
perturative scheme for large enough average fracture openings for them to
be accurate. 

Future work will be devoted to the study the relationship of
between $\Delta$ and the statistics of
the aperture fields.

The worst path method is accurate even if the aperture field shows structural
anisotropy. Such situation is achieved by laterally displacing the
fracture walls leading, as observed on natural fractures, to an
anisotropic permeability field. The proposed method can be further
extended to other transport properties such as diffusion or
electrical conductivity. Different statistical fields such as log
normal permeability fields which also give rise to heterogenous flow
structures will also be of interest as well
as three-dimensional permeability fields.

\section{acknowledgments}
We thank J.\ P.\ Hulin for important comments and discussion. A.H.\ thanks
Laboratoire FAST for hospitality during his stay at Orsay and the
Universit{\'e} de Paris 11 for funding.  This work has been greatly
facilitated by the GdR 2990 HTHS and the {\it Triangle de la Physique.\/}

\end{document}